\documentclass[preprintnumbers,showpacs,amsmath,amssymb,floatfix,prd,onecolumn,superscriptaddress,nofootinbib]{revtex4}
\usepackage{bbm}
\usepackage{graphicx}
\usepackage{epsfig}
\usepackage{bm}
\usepackage{amsfonts}
\usepackage{epstopdf}

\begin{document}

\title{The Casimir force of Quantum Spring in the (D+1)-dimensional spacetime}

\author{Xiang-hua Zhai}\email{zhaixh@shnu.edu.cn}

\author{Xin-zhou Li}\email{kychz@shnu.edu.cn}

\author{Chao-Jun Feng}
\email{fengcj@shnu.edu.cn}

\affiliation{Shanghai United Center for Astrophysics (SUCA), \\ Shanghai Normal University,
    100 Guilin Road, Shanghai 200234, China}

\begin{abstract}
The Casimir effect for a massless scalar field on the helix boundary
condition which is named as quantum spring is studied in our recent
paper\cite{Feng}. In this paper, the Casimir effect of the quantum spring is investigated in $(D+1)$-dimensional spacetime for the massless and massive scalar fields by
using the zeta function techniques. We obtain the exact results of
the Casimir energy and Casimir force for any $D$, which indicate
a $Z_2$ symmetry of the two space dimensions. The Casimir energy
and Casimir force have different expressions for odd and even
dimensional space in the massless case but in both cases the force is attractive. In the case of odd-dimensional space, the Casimir energy density can be expressed by the Bernoulli numbers, while in the even case it can be expressed by the $\zeta$-function. And we
also show that the Casimir force has a maximum value which depends on the spacetime dimensions. 
In particular, for a massive scalar field, we found that the Casimir force varies as the mass of the field changes.
\end{abstract}

 \maketitle


\section{Introduction}\label{sec:intro}
The work done by Casimir\cite{Casimir:1948dh} more than 60 years ago
starts an important research field as one of the direct
manifestations of the existence of the zero-point vacuum
oscillations. The Casimir effect has now been extensively studied
both in experiment and theory, especially in the past decade due to
the precise verification benefited from the applications of the
modern laboratory techniques\cite{klimchitskaya}. And it is still a
topic full of life that attracted increasing interests in both
fundamental and applied science\cite{Plunien:1986ca}. The Casimir
effect, in its simplest form, is the attraction between two
plane-parallel uncharged perfectly conducting plates in vacuum.

The nature of the Casimir force may depend on (i) the background
field, (ii) the spacetime dimensionality, (iii) the type of boundary
conditions, (iv) the topology of spacetime, (v) the finite
temperature. The most evident example of the dependence on the
geometry is given by the Casimir effect inside a rectangular box
\cite{Plunien:1986ca, Lukosz}. The detailed calculation of the
Casimir force inside a D-dimensional rectangular cavity was shown in
\cite{Li}, in which the sign of the Casimir energy depends on the
length of the sides. The Casimir force arises not only in the
presence of material boundaries, but also in spaces with nontrivial
topology. For example, we get the scalar field on a flat manifold
with topology of a circle $S^1$. The topology of $S^1$ causes the
periodicity condition $\phi(t,0)=\phi(t,C)$, where $C$ is the
circumference of $S^1$, imposed on the wave function which is of the
same kind as those due to boundary. Similarly, the antiperiodic
conditions can be drawn on a M\"obius strip. The $\zeta$-function
regularization procedure is a very powerful and elegant technique
for the Casimir effect. Rigorous extension of the proof of Epstein
$\zeta$-function regularization has been discussed in
\cite{Elizalde}. Vacuum polarization in the background of on string
was first considered in \cite{Helliwell:1986hs}. The generalized
$\zeta$-function has many interesting applications, e.g., in the
piecewise string \cite{Li:1990bz}. Similar analysis has been applied
to monopoles \cite{BezerradeMello:1999ge}, p-branes
\cite{Shi:1991qc} or pistons \cite{Zhai}.Recently,the Casimir effect
has been paid more attention due to the development of precise
measurements \cite{Decca:2007yb}, and it has been applied to the
fabrication of microelectromechanical systems (MEMS)\cite{MEMS}.
Furthermore, some new methods have developed for computing the
Casimir energy between a finite number of compact objects
\cite{Emig:2007cf}. In our recent paper, the Casimir effect for a massless scalar field on
the helix boundary condition is investigated by using the zeta function techniques\cite{Feng}. We find
that the Casimir force is very much like the force on a spring that
obeys the hooke's law in mechanics. However, in this case, the force
comes from a quantum effect, and so we would like to call this
structures as a quantum spring\cite{Feng}. On the other hand, the Casimir effect for the massive scalar field is also studied by some authors\cite{Bordag2}. As is known that the Casimir effect disappears as the mass of the field goes to infinity since there are no more quantum fluctuations in the limit. But the precise way the Casimir energy varies as the mass changes is worth studying\cite{Plunien:1986ca}.

In this paper, we study the quantum spring in ($D+1$)-dimensional
spacetime. We obtain the exact results of the Casimir energy and
Casimir force for the massless and massive scalar fields in the
($D+1$)-dimensional spacetime. The final results also tell us that
there is a $Z_2$ symmetry of the two space dimensions. And we also
show that the Casimir force has a maximum value which depends on the
spacetime dimensions for both massless and massive cases.
Especially, we show that the Casmir force varies as the mass of the
field changes.

\section{Topology of the flat ($D$+1)-dimensional spacetime}\label{sec:casimir}

As mentioned in Section \ref{sec:intro}, the Casimir effect arises
not only in the presence of material boundaries, but also in spaces
with nontrivial topology. For example, we get the scalar field on a
flat manifold with topology of a circle $S^1$. The topology of $S^1$
causes the periodicity condition $\phi(t,0)=\phi(t,C)$. Before we
consider complicated cases in the flat spacetime, we have to discuss
the lattices.

A lattice $\Lambda$ is defined as a set of points in a flat
($D+1$)-dimensional spacetime $\mathcal{M}^{D+1}$, of the form
\begin{equation}\label{lattice}
    \Lambda = \left\{ ~ \sum_{i=0}^{D} n_i \mathbbm{e}_i ~|~ n_i \in \mathcal{Z} ~\right\} \,,
\end{equation}
where $\{\mathbbm{e}_i\}$ is a set of basis vectors of $\mathcal{M}^{D+1}$. In terms of the components $v^i$ of vectors
$\mathbb{V} \in \mathcal{M}^{D+1} $, we define the inner products as
\begin{equation}\label{inner prod}
    \mathbb{V} \cdot \mathbb{W} = \epsilon(a)v^iw^j\delta_{ij} \,,
\end{equation}
with $\epsilon(a)=1$ for $i=0$, $\epsilon(a)=-1$ for otherwise. In the $x^1-x^2$ plane, the sublattice
$\Lambda''\subset\Lambda'\subset\Lambda$ are
\begin{equation}\label{sub1}
   \Lambda' = \left\{ ~  n_1 \mathbbm{e}_1 + n_2 \mathbbm{e}_2 ~|~ n_{1,2} \in \mathcal{Z} ~\right\} \,,
\end{equation}
and
\begin{equation}\label{sub2}
   \Lambda'' = \left\{ ~  n(\mathbbm{e}_1 + \mathbbm{e}_2) ~|~ n \in \mathcal{Z} ~\right\} \,.
\end{equation}

The unit cylinder-cell is the set of points
\begin{eqnarray}
 \nonumber
   U_c &=& \bigg\{\mathbb{X} = \sum_{i=0}^{D}x^i \mathbbm{e}_i ~|~ 0\leq x^1 < a,
 -h\leq x^2 < 0 , \\ && -\infty <x^0<\infty, -\frac{L}{2} \leq x^T\leq \frac{L}{2}\bigg\} \,,\label{cell}
\end{eqnarray}
where $T = 3,\cdots, D$. When $L\rightarrow\infty$, it contains precisely one lattice point (i.e. $\mathbb{X} = 0$),
and any vector $\mathbb{V}$ has precisely one "image" in the unit cylinder-cell, obtained by adding a sublattice vector
to it.

In this paper, we choose a topology of the flat ($D+1$)-dimensional
spacetime: $U_c\equiv U_c +\mathbbm{u}, \mathbbm{u} \in \Lambda''$.
This topology causes the helix boundary condition for a Hermitian
massless or massive scalar field
\begin{equation}\label{helxi boundary condition}
   \phi(t, x^1 + a, x^2, x^T) =  \phi(t, x^1 , x^2+h, x^T) \,,
\end{equation}
where, if $a=0$ or $h=0$, it returns to the periodicity boundary condition.

\section{Massless scalar field }

\subsection{Casmir energy in the massless case}

In calculations on the Casimir effect, extensive use is made of
eigenfunctions and eigenvalues of the corresponding field equation.
A Hermitian massless scalar field $\phi(t, x^\alpha, x^T)$ defined
in the ($D+1$)-dimensional flat spacetime satisfies the free
Klein-Gordon equation:
\begin{equation}\label{eom}
    \left(\partial_t^2 - \partial_i^2\right)\phi(t, x^\alpha, x^T) = 0 \,,
\end{equation}
where $i=1,\cdots, D; \alpha=1,2; T=3,\cdots, D$. Under the boundary condition (\ref{helxi boundary condition}), the
modes of the field are then
\begin{equation}\label{modes}
    \phi_{n}(t, x^\alpha, x^T)= \mathcal{N} e^{-i\omega_nt+ik_x x+ik_z z + ik_Tx^T }\,,
\end{equation}
where $\mathcal{N}$ is a normalization factor and $x^1=x, x^2=z$, and we have
\begin{equation}\label{energy}
    \omega_n^2 = k_{T}^2 + k_x^2 + \left( -\frac{2\pi n}{h}+\frac{k_x}{h}a \right)^2 = k_{T}^2 + k_z^2 + \left( \frac{2\pi n}{a}+\frac{k_z}{a}h
    \right)^2 \,.
\end{equation}
Here, $k_x$ and $k_z$ satisfy
\begin{equation}\label{kxkz}
    a k_x - hk_z = 2n\pi\,, (n=0,\pm1,\pm2,\cdots) \,.
\end{equation}
In the ground state (vacuum), each of these modes contributes an
energy of $\omega_n/2$. The energy density of the field in
($D+1$)-dimensional spacetime is thus given by
\begin{widetext}
\begin{eqnarray}
\nonumber
  &E^{D}& = \frac{1}{2 a}
  \int \frac{d^{D-1}k}{(2\pi)^{D-1}} \sum_{n=-\infty}^{\infty} \sqrt{k_T^2 + k_z^2 + \left( \frac{2\pi n}{a}+\frac{k_z}{a}h
    \right)^2  } \,, \\&&\label{tot energy}
\end{eqnarray}
\end{widetext}
\noindent where we have assumed $a\neq 0$ without losing generalities. Eq.(\ref{tot energy}) can be rewritten as
\begin{eqnarray}
\nonumber
  &E^{D}& = \frac{1}{2 a \sqrt{\gamma}}
  \int \frac{d^{D-1}u}{(2\pi)^{D-1}} \sum_{n=-\infty}^{\infty} \sqrt{u^2 + \left( \frac{2\pi n}{a \gamma}\right)^2  } \,, \\&&\label{re-tot energy}
\end{eqnarray}
where we have defined
\begin{equation}\label{gamma}
\gamma \equiv 1+ \frac {h^2}{a^2}.
\end{equation}

Using the mathematical identity,
\begin{equation}
\int_{-\infty}^{\infty} f(u) d^{D-1} u=\frac {2\pi^{\frac
{D-1}{2}}}{\Gamma \left (\frac{D-1}{2} \right )}\int_{0}^{\infty}
u^{D-2} f(u) du,
\end{equation}
we get
\begin{equation}
\int d^{D-1} u \sqrt{u^2+\left (\frac{2\pi n}{a \sqrt{\gamma}} \right )^2}=\frac{\pi ^{\frac {D-1}{2}}\Gamma \left (-\frac D 2 \right )}{\Gamma \left (-\frac 1 2 \right )}\left (\frac{2 \pi n}{a \sqrt{\gamma}} \right )^D \,.
\end{equation}

\noindent thus the energy density in Eq.(\ref{tot energy}) is
reduced to
\begin{equation}\label{fienergy}
E^{D}=-\frac {\pi^{\frac D 2}}{a^{D+1}\gamma^{\frac {D+1}{2}}}\Gamma \left (-\frac D 2 \right )\zeta(-D) \,,
\end{equation}

\noindent where $\zeta(-D)$ is the Riemann $\zeta$ function.

In the case of $D=2j+1$, Eq.(\ref{fienergy}) shows directly the
Casimir energy and the final expression is
\begin{equation}
E_R ^{2j+1}=-\frac{(2\pi)^{j+1}|B_{2j+2}|}{(2j+1)!!(2j+2)(a^2+h^2)^{j+1}} \,,
\end{equation}
\noindent where $j=1,2,...$ and the Bernoulli numbers are $B_2=\frac
1 6,B_4=-\frac 1 {30},B_6=\frac 1{42},B_8=-\frac 1 {30},B_{10}=\frac
5 {66},B_{12}=-\frac {691}{2730},B_{14}=\frac 7
6,B_{16}=-\frac{3617}{510},...$. In the case of $D=2j$, because
$\Gamma(-j)$ is a pole of order 1, Eq.(\ref{fienergy}) should be
regularized by using the reflection relation
\begin{equation}\label{rel}
    \Gamma\left(\frac{s}{2}\right)\zeta(s) = \pi^{s-\frac{1}{2}} \Gamma\left(\frac{1-s}{2}\right)\zeta(1-s)\,.
\end{equation}

\noindent Then, the final expression is
\begin{equation}
	E_R ^{2j}=-\frac{(2j-1)!!\zeta(2j+1)}{(2\pi)^j(a^2+h^2)^{j+\frac 1 2}} \,.
\end{equation}

\noindent Obviously, we have the symmetry of $a\leftrightarrow h$ in
both cases. It is worth noting that the Casimir energy has different
expressions for the odd and even space dimensions.

\subsection{The Casimir force in the massless case}\label{sec:force}

The Casimir force on the $x$ direction is given by
\begin{equation}
F_a=-\frac {\partial E_R^{D+1}}{\partial a} \,.
\end{equation}

For the odd-dimensional space, the final expression for the Casimir force is
\begin{equation}
F_a^{2j+1}=-\frac{(2\pi)^{j+1}|B_{2j+2}|a}{(2j+1)!!(a^2+h^2)^{j+2}},
\end{equation}

\noindent which has a maximum value of magnitude

\begin{equation}
F_{a,max}^{2j+1}=-\frac{(2\pi)^{j+1}|B_{2j+2}|}{(2j+1)!!h^{2j+3}}\sqrt{\frac{(2j+3)^{2j+3}}{(2j+4)^{2j+4}}}
\end{equation}

\noindent at $a=\frac {h}{\sqrt{2j+3}}$. While for the even-dimensional space, the final expression for the Casimir force is
\begin{equation}
F_a^{2j}=-\frac{2(j+\frac 1 2)(2j-1)!!\zeta(2j+1)a}{(2\pi)^j(a^2+h^2)^{j+\frac 3 2}},
\end{equation}

\noindent and the maximum value of force magnitude

\begin{equation}
F_{a,max}^{2j}=-\frac{2(j+\frac 1 2)(2j+1)!!\zeta(2j+1)}{(2\pi)^j h^{2j+2}}\sqrt{\frac{(2j+2)^{2j+2}}{(2j+3)^{2j+3}}}
\end{equation}

\noindent is obtained at $a=\frac {h}{\sqrt{2j+2}}$. The force in both cases is attractive force. The results for $F_h$ are similar to those of $F_a$ because of the symmetry between $a$ and $h$. We list the Casimir energy and forces in the two
directions in Table~\ref{Table:d23} for $D=2,3,4,5$.
\begin{table}
\centering
{\begin{tabular}{@{}cccc@{}}
\toprule $D$ & $E_R^D$ & $F_a^D$ & $F_h^D$
\\\colrule 2\hphantom{00}&\hphantom{0}$-\frac{\zeta(3)}{2\pi }\frac{1}{(a^2+h^2)^{\frac 3 2}}$
		\hphantom{00}&\hphantom{0}$-\frac{3 \zeta(3)}{2\pi }\frac{a}{(a^2+h^2)^{-\frac 5 2}}$
		\hphantom{00}&\hphantom{0}$-\frac{3 \zeta(3)}{2\pi }\frac{h}{(a^2+h^2)^{-\frac 5 2}}$\\ 
	      3\hphantom{00}&\hphantom{0}$-\frac{\pi ^2}{90 }\frac{1}{(a^2+h^2)^2}$
	        \hphantom{00}&\hphantom{0}$-\frac{2\pi ^2}{45 }\frac{a}{(a^2+h^2)^3}$
	        \hphantom{00}&\hphantom{0}$-\frac{2\pi ^2}{45 }\frac{h}{(a^2+h^2)^3}$\\
	     4\hphantom{00}&\hphantom{0}$-\frac{3\zeta(5)}{4\pi ^2 }\frac{1}{(a^2+h^2)^{\frac 5 2}}$
	       \hphantom{00}&\hphantom{0}$-\frac{15 \zeta(5)}{4\pi ^2 }\frac{a}{(a^2+h^2)^{-\frac 7 2}}$
	       \hphantom{00}&\hphantom{0}$-\frac{15 \zeta(5)}{4\pi ^2 }\frac{h}{(a^2+h^2)^{-\frac 7 2}}$\\
	     5\hphantom{00}&\hphantom{0}$-\frac{2\pi ^3}{945}\frac{1}{(a^2+h^2)^3}$
	       \hphantom{00}&\hphantom{0}$-\frac{4 \pi ^3}{315}\frac{a}{(a^2+h^2)^4}$
	       \hphantom{00}&\hphantom{0}$-\frac{4 \pi ^3}{315}\frac{h}{(a^2+h^2)^4}$\\
\botrule
\end{tabular}}
\caption{\label{Table:d23}The Casimir energy and forces for $D=2,3,4,5$.}
\end{table}
In Fig. 1, we  illustrate  the behavior of the Casimir force on
$x$ direction in $D=3$ dimension. The curves from the bottom to top
correspond to $h=0.9,1.0,1.1,1.2$ respectively. It is clearly seen that the Casimir force decreases with $h$ increasing and the maximum value of the force magnitude
$\frac{2\pi^2}{45h^5}\sqrt{\frac{5^5}{6^6}}$ appears at
$a=\frac{h}{\sqrt{5}}$. And in Fig. 2 we  illustrate  the behavior of the Casimir force on
$x$ direction in different dimensions. The curves from the bottom to
top correspond to $D=2,3,4,5$ respectively. We take $h=1.5$ in this
figure. It is clearly seen that the Casimir force decreases with $D$
increasing, and the value of $a$ where the maximum value of the
force is achieved also gets smaller with $D$ increasing.

\begin{figure}[h]
\begin{center}
\includegraphics[width=0.4\textwidth]{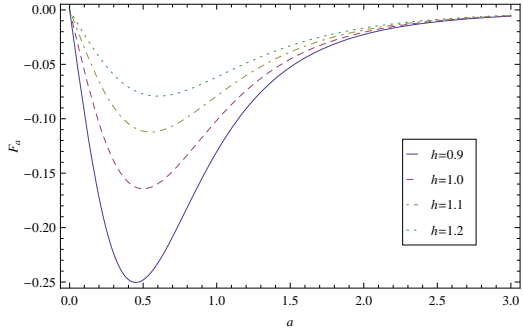}
\caption{The Casimir force on the $x$ direction \textit{vs.}$a$ in
$D=3$ dimension for different $h$. The Casimir force decreases with
$h$ increasing and the maximum value of the force magnitude
$\frac{2\pi^2}{45h^5}\sqrt{\frac{5^5}{6^6}}$ appears at
$a=\frac{h}{\sqrt{5}}$.}
\end{center}
\end{figure}

\begin{figure}[h]
\begin{center}
\includegraphics[width=0.4\textwidth]{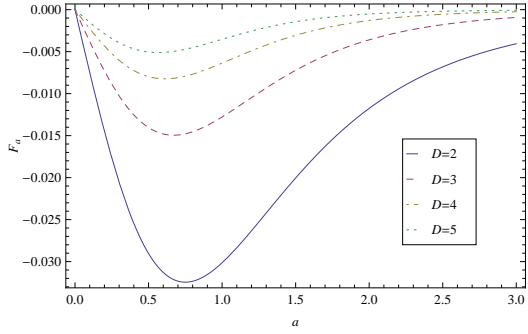}
\caption{The Casimir force on the $x$ direction \textit{vs.}$a$ in
different dimensions. Here we take $h=1.5$. It is clearly seen that
the Casimir force decreases with $D$ increasing, and the value of
$a$ where the maximum value of the force is achieved also gets
smaller with $D$ increasing.}
\end{center}
\end{figure}

\section{Massive scalar field}

In this section, we extend our discussion to the case of massive scalar field. A scalar field $\phi$ defined in the $(D+1)$-dimensional flat spacetime satisfies the equation as follows

\begin{equation}
    \left(\partial_t^2 - \partial_i^2+\mu^2\right)\phi(t, x^\alpha, x^T) = 0 \,,
\end{equation}

\noindent where $\mu$ is the mass of $\phi$ field. In this case, we have
\begin{eqnarray}\label{frequency}
    \omega_n^2 &=& k_{T}^2 + k_x^2 + \left( -\frac{2\pi n}{h}+\frac{k_x}{h}a \right)^2+\mu^2\nonumber\\
     &=& k_{T}^2 + k_z^2 + \left( \frac{2\pi n}{a}+\frac{k_z}{a}h
    \right)^2+\mu^2 \,.
\end{eqnarray}
\noindent where $k_x$ and $k_z$ satisfy Eq.(10). The Casimir energy density of the massive scalar field in the ($D+1$)-dimensional spacetime is thus given by

\begin{widetext}
\begin{eqnarray}
\nonumber
  &E_{\mu}^{D}& = \frac{1}{2 a}
  \int \frac{d^{D-1}k}{(2\pi)^{D-1}} \sum_{n=-\infty}^{\infty} \sqrt{k_T^2 + k_z^2 + \left( \frac{2\pi n}{a}+\frac{k_z}{a}h
    \right)^2 +\mu^2 } \,, \\&&\label{tot energy-m}
\end{eqnarray}
\end{widetext}

To regularize Eq.(27), we use the functional relation
\begin{widetext}
\begin{equation}
\sum_{n=-\infty}^{\infty} \left (bn^2+\mu^2\right )^{-s}=\frac{\sqrt{\pi}}{\sqrt{b}}\frac{\Gamma\left (s-\frac 12\right )}{\Gamma(s)}\mu^{1-2s}
+\frac{\pi^s}{\sqrt{b}}\frac{2}{\Gamma(s)}\sum_{n=-\infty}^{\infty\hspace{0.2cm}\prime}\mu^{\frac 12-s}\left (\frac{n}{\sqrt{b}}\right )^{s-\frac 12}K_{\frac 12-s}\left( 2\pi \mu \frac{n}{\sqrt{b}}\right )
\end{equation}
\end{widetext}

\noindent where $K_{\nu}(z)$ is the modified Bessel function and the
prime means that the term $n=0$ has to be excluded. After tedious
deduction, we have
\begin{widetext}
\begin{equation}
E_{R,\mu}^D=-\frac{\mu^{D+1}\Gamma\left (-\frac{D+1}{2}\right
)}{2^{D+2}\pi^{\frac{D+1}{2}}}-2\left
(\frac{\mu}{2\pi\sqrt{a^2+h^2}}\right
)^{\frac{D+1}{2}}\sum_{n=1}^{\infty}n^{-\frac{D+1}{2}}K_{\frac{D+1}{2}}\left
( n\mu \sqrt{a^2+h^2}\right )
\end{equation}
\end{widetext}

\noindent For $\nu>0$ and $z\rightarrow 0$, the Bessel function has
the asymptotic expression $K_{\nu}(z)\rightarrow
\frac{2^{\nu-1}\Gamma(\nu)}{z^{\nu}}$, so it is not difficult to
find that when $\mu\rightarrow 0$, the Casimir energy recover the
result of the massless case.

 Using Eq.(20) and
$K_{\nu}^{\prime}(z)=\frac{\nu}{z}K_{\nu}(z)-K_{\nu+1}(z)$ where
$K_{\nu}^{\prime}(z)=dK_{\nu}(z)/dz$, we have the Casimir force
\begin{widetext}
\begin{equation}
F_{a,\mu}=-\frac{2\mu a\left ( (\mu a)^2+(\mu h)^2\right
)^{\frac{D+1}{4}}}{\left ( 2\pi\right )^{\frac{D+1}{2}}\left (
a^2+h^2\right )^{\frac
{D+2}{2}}}\sum_{n=1}^{\infty}n^{-\frac{D-1}{2}}K_{\frac{D+3}{2}}\left
( n\mu \sqrt{a^2+h^2}\right )
\end{equation}
\end{widetext}

We study numerically the behavior of the Casimir force on $x$
direction as a function of $a$ for different $h$ and $D$. We find
that the Casimir force is still attractive and it has a maximum
value similarly to massless case. For given values of $D$ and $\mu$,
the behavior of the force for different $h$ is similar to that in
massless case. But for given values of $h$ and $\mu$, the behavior
of the force for different $D$ is opposite to that in massless case.
The force increases with $D$ increasing and the position of the
maximum value move to larger $a$ as $D$ increasing. We plot the
force as a function of $a$ in Fig. 3 for $\mu=1,h=1$ and $D=2,3,4,5$
respectively and it is easy to find the difference between Fig. 3
and Fig. 2.

Because the precise way the Casimir force varies as the mass changes is worth studying, we give the rate of massive and massless cases as follows
\begin{widetext}
\begin{equation}
\frac{F_{a,\mu}}{F_{a,0}}=\frac{\left ( (\mu a)^2+(\mu h)^2\right
)^{\frac{D+3}{4}}}{(D+1)2^{\frac {D-1}{2}}\Gamma\left ( \frac
{D+1}{2}\right
)\zeta(D+1)}\sum_{n=1}^{\infty}n^{-\frac{D-1}{2}}K_{\frac{D+3}{2}}\left
( n\mu \sqrt{a^2+h^2}\right )
\end{equation}
\end{widetext}

In the case of odd-dimensional space, Eq.(31) can be reduced to
\begin{widetext}
\begin{equation}
\frac{F_{a,\mu}}{F_{a,0}}=\frac{2(2j+1)!!\left ( (\mu a)^2+(\mu
h)^2\right )^{\frac{j+2}{2}}}{\left (2\pi\right
)^{2j+2}|B_{2j+2}|}\sum_{n=1}^{\infty}n^{-j}K_{j+2}\left ( n\mu
\sqrt{a^2+h^2}\right )
\end{equation}
\end{widetext}
\noindent where $j=1,2,...$. Obviously, the ratio tends to 1 when $\mu\rightarrow 0$ and it tends to zero when $\mu\rightarrow \infty$.

Fig.4 is the illustration of the ratio of the Casimir force in
massive case to that in massless case varying with the mass in $D=3$
dimension. The curves correspond to $a=1$ and $h=0.1,1,2,3$
respectively. Fig.5 is the illustration of the ratio of the Casimir
force in massive case to that in massless case varying with the mass
for different dimensions. The curves correspond to $a=1, h=0.1$ and
$D=2,3,4,5$ respectively. It is clearly seen from the two figures
that the Casimir force decreases with $\mu$ increasing, and it
approaches zero when $\mu$ tends to infinity. The plots also tell us
the Casimir force for a massive field decreases with $h$ increasing
but it increases with $D$ increasing. For the latter, the behavior
of the Casimir force in massive case is different from that of
massless case.

\begin{figure}[h]
\begin{center}
\includegraphics[width=0.4\textwidth]{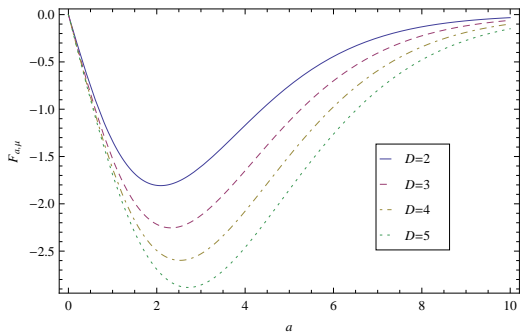}
\caption{The Casimir force on the $x$ direction \textit{vs.}$a$ in
different dimensions for a massive scalar field. Here we take
$h=1,\mu=1$ and $D=2,3,4,5$ respectively. It is clearly seen that
the Casimir force increases with $D$ increasing, and the maximum
value of the force moves to larger $a$ as as $D$ increasing, which
is a feature that opposite to massless case.}
\end{center}
\end{figure}

\begin{figure}[h]
\begin{center}
\includegraphics[width=0.4\textwidth]{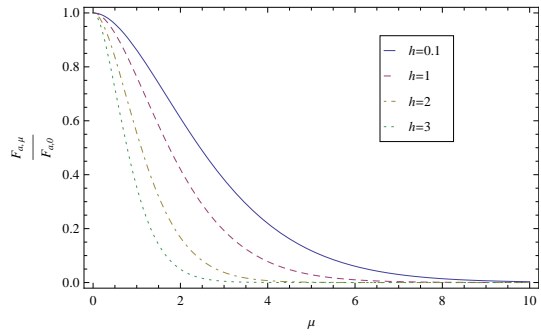}
\caption{the ratio of the Casimir force in massive case to that in massless case varying with the mass for different $h$ in $D=3$ dimension. The curves correspond to $a=1$ and $h=0.1,1,2,3$ respectively.}
\end{center}
\end{figure}

\begin{figure}[h]
\begin{center}
\includegraphics[width=0.4\textwidth]{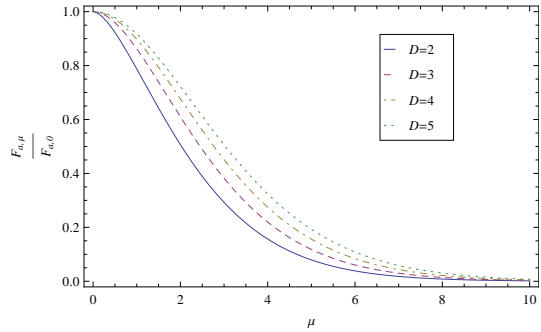}
\caption{the ratio of the Casimir force in massive case to that in massless case varying with the mass for different dimensions. We take $a=1$ and $h=0.1$. }
\end{center}
\end{figure}

\section{Conclusion }

In this paper, the quantum spring is investigated in $(D+1)$-dimensional spacetime by using the zeta function techniques for both massless and massive scalar field, and  in conclusion, we summarize our results as the following
\begin{itemize}
\item For the massless scalar field, the exact expressions of the Casimir energy and Casimir force are obtained in arbitrary $D+1$ dimensional spacetime, and when $D$ is odd, the energy and force could be expressed in terms of the Bernoulli numbers, but for even values of $D$, they can only be expressed in terms of the Riemann $\zeta$ function.  
\item For the massive scalar field, we also get the exact results of the Casimir energy and force. To  see the effect of the mass, we  compare  the results with that of the massless one and we found that the Casimir force approaches the result of the force in the massless case when the mass tends to zero and vanishes when the mass tends to infinity.
\item For both massless and massive scalar field,  the Casimir force in $x$ direction decreases when $h$ is increasing and there is a symmetry
 of $a\leftrightarrow h$ for the Casimir energy. The Casimir force has a maximum value, and the critical value of $a$ to get this maximum value increases with $h$ increasing in both cases. 
\item There is a little different behavior of the Casimir force for the massless and massive field. That is the Casimir force and its maximum critical value of $a$  decrease with $D$ increasing in the massless case, but increase with $D$ increasing in the massive case.
\end{itemize}

  As is known that the Casimir effect can apply to the cosmology
  with extra dimensions, the effect of the quantum spring in the
  (1+3+2)- dimensional cosmology is worth considering and we will study
  it in our further work\cite{Zhai2}.

\acknowledgments

This work is supported by National Nature Science Foundation of China under Grant No. 10671128, National Education Foundation of China grant No. 2009312711004 and Shanghai Natural Science
Foundation, China grant No. 10ZR1422000.

\end{document}